\documentclass[10pt, conference, letterpaper]{IEEEtran}
\ifCLASSINFOpdf
\else
\fi
\usepackage{graphicx}
\usepackage{epstopdf}
\DeclareGraphicsExtensions{.eps}
\graphicspath{ {images/} }

\usepackage{amsmath}

\usepackage{amsmath}

\usepackage{amsthm}

\theoremstyle{definition}

\begin{document}
%
\title{Spatial Modulation-Spatial Multiplexing in \\ Massive MIMO}
\author{\IEEEauthorblockN{Garimella Rama Murthy and Kunal Sankhe}
\IEEEauthorblockA{Signal Processing and Communication Research Center\\
International Institute of Information Technology\\
Hyderabad, India\\
Email: rammurthy@iiit.ac.in and kunal.sankhe@research.iiit.ac.in}
}


%


\maketitle




%
\IEEEpeerreviewmaketitle

\section{Introduction}

Massive MIMO, a candidate for 5G technology, promises significant gains in wireless data rates and link reliability by using large numbers of antennas (more than 64) at the base transceiver station (BTS). 
Extra antennas help by focusing the transmission and reception of signal energy into ever-smaller regions of space. This brings huge improvements in throughput. However, it requires a large number of Radio Frequency (RF) chains (usually equal to number of transmit antennas), which is a major drawback.

One approach to overcome these issues is to use Spatial Modulation (SM) \cite{SM-Basic}. SM is a recently developed transmission technique that uses multiple antennas \cite{SM-Basic}. In SM, an index of transmit antenna is used as an additional source of information to improve the overall spectral efficiency. In particular, a group of any number of information bits is mapped into two constellations: a signal constellation based on modulation scheme and a spatial constellation to encode the index of the transmit antenna. At any time instant, only one transmit antenna is active, whereas other transmit antennas radiate zero power. This completely avoids inter channel interference at the receiver and relaxes the stringent requirement of synchronization among the transmit antennas. In addition, unlike conventional MIMO system, SM system does not require multiple RF chains at the transmitter. At the receiver, a low complexity decoder such as maximum receive ratio combining (MRRC) is used to estimate the index of active transmit antenna, and after which transmitted symbol is estimated. Using these two estimates, a spatial demodulator retrieves the group of information bits.

However, a low spectral efficiency is main drawback of SM. Therefore, a combination of SM with Spatial Multiplexing is an effective way to increase spectral efficiency with limited number of RF chains. 

\section{Spatial Modulation-Spatial Multiplexing MIMO}


\begin{figure}[h]
	\centering
	\includegraphics[width=3.2in]{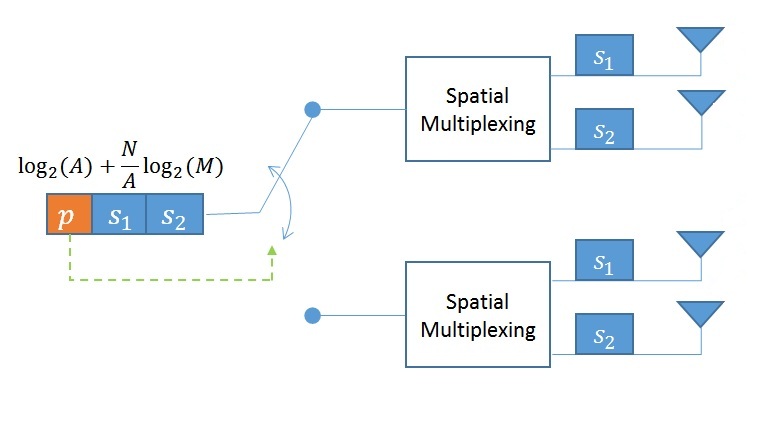}
	\label{fig:Illustration}
	\end{figure}

Let $b$ be the $ \log_2(K) + \frac{N}{K}\log_{2}(M)$ bits emitted by the source, where $K$ is the number of 
number of available RF chains and N is the total numbet of antennas. The information bits are divided into two blocks of length $\log_{2}(K)$ and $\frac{N}{K}\log_{2}(M)$. 

The first group of $\log_{2}(K)$ information bits determine which group of antennas is activated out of $\frac{N}{K}$ antenna groups. Then $ K\log_{2}(M)$ bits are divided further into K groups of $\log_{2}(M)$ bits. Now, each $\log_{2}(M)$ bits are modulated using M-QAM modulation and simultaneously transmitted over $K$ antennas using Spatial Multiplexing mode of MIMO.    

To illustrate, consider an example, where we have $N=4 $ transmit antennas and $K=2$ RF chains. Now, a set of $N$ transmit antennas are divided into $\frac{N}{K}=\frac{4}{4}=2$ groups, while each group contains $K=2$ antennas. A spectral efficiency of 5 bits/symbol/Hz ($\eta = \log_2(2) + 2\log_{2}(4)=5$) bits/symbol/Hz is considered for this example, with $M=4$ as constellation of QAM modulation.  
Now, the information bits are divided into two parts: the first part of data, i.e., ($\log_2(2)=1$) bit determine which group of antennas will remain active, and the other part data $2\log_{2}(4)=4$ bits is further divided into two blocks of length ($\log_2(M)=2$) bits. Two bits of each block is  modulated using $M=4$ modulation and simultaneously transmitted from antennas. 

The advantage is if we use only Spatial modulation, the spectral efficiency would be $\log_2(4)+\log_2(4)=4$ bits/s/Hz, whereas this configuration will have spectral efficiency of $\log_2(2)+2\log_2(4)=5$ bits/s/Hz with little more complexity. In addition, the number of RF chains required is only two. 

In future, the low complexity detection algorithm will be developed.

\begin{thebibliography}{9}

\bibitem{SM-Basic} 
M. Renzo, H. Haas, A. Ghrayeb, S. Sugiura, and L. Hanzo, "Spatial modulation for generalized MIMO: Challenges, opportunities, and implementation," Proceedings of the IEEE 102.1 (2014): 56-103.
\end{thebibliography}
\end{document}